\documentclass[fleqn,10pt]{wlscirep}
\usepackage[utf8]{inputenc}
\usepackage[T1]{fontenc}
\usepackage{bm}
\usepackage{my_aas_macros}
\usepackage{tcolorbox}
\usepackage{wrapfig}

\newcommand{\msun}{\,{\rm M_\odot}}

\title{The origins of massive black holes}

\author[1,*]{Marta Volonteri}
\author[2,3]{M\'elanie Habouzit}
\author[4,5]{Monica Colpi}
\affil[1]{Institut d’Astrophysique de Paris, Sorbonne Universit\'e, CNRS, UMR 7095, 98 bis bd Arago, 75014 Paris, France}
\affil[2]{Max-Planck-Institut f\"ur Astronomie, K\"onigstuhl 17, D-69117 Heidelberg, Germany}
\affil[3]{Zentrum f\"ur Astronomie der Universit\"at Heidelberg, ITA, Albert-Ueberle-Str. 2, D-69120 Heidelberg, Germany}
\affil[4]{Dipartimento di Fisica G. Occhialini, Universita di Milano-Bicocca, Piazza della Scienza 3, I-20126 Milano, Italy}
\affil[5]{INFN, Sezione di Milano-Bicocca, Piazza della Scienza 3, I-20126 Milano, Italy}

\affil[*]{e-mail: martav@iap.fr}

\begin{abstract}
Massive black holes (MBHs) inhabit galaxy centers, power luminous quasars and  Active Galactic Nuclei (AGN) and shape their cosmic environment with the energy they produce. 
The origins of MBHs remain a mystery and the recent detection by LIGO/Virgo of an almost 150 solar mass black hole has revitalized the question of whether there is a continuum between ``stellar'' and ``massive'' black holes  and what the {\it seeds} of MBHs are. Seeds could have formed in the first galaxies, or could be also related to the collapse of horizon-sized regions in the early Universe. 
Understanding the origins of MBHs straddles fundamental physics, cosmology and astrophysics  and it bridges the fields of gravitational wave physics and traditional astronomy. 
With several facilities in the next 10-15 years we foresee the possibility of discovering MBHs' avenues of formation. 
In this article we link three main topics: the channels of black hole seed formation, the journey from seeds to massive black holes, the diagnostics on the origins of MBHs. We highlight and critically discuss current unsolved problems and touch on recent developments that stirred the community. 
\end{abstract}

\begin{document}

\flushbottom
\maketitle

\thispagestyle{empty}

Observations reveal that astrophysical black holes come in two different families: stellar black holes, relics of massive stars and widespread in all galaxies of the Universe, and massive black holes (MBHs) at the centers of most of the galaxies in today's Universe, with masses up to and above several billions of solar masses. MBHs fill the upper end of the mass spectrum of black holes, which at the lower-mass end is populated by stellar black holes. The lightest MBH candidate, in the galaxy NGC 205 \cite{2019ApJ...872..104N}, weighs $6800\msun$, while the heaviest stellar black hole, GW190521 \cite{2020ApJ...900L..13A} resulting from a merger, has mass $\sim 142  \msun$.
MBHs are central actors in the ecology of galaxy formation through the energy produced as they grow and shine as quasars \cite{2012ARA&A..50..455F}, and there is a sheer difficulty in explaining in a single picture the presence of quasars powered by billion solar mass MBHs when the Universe was less than a billion years old \cite{2001AJ....122.2833F}, the ubiquity of MBHs in galaxies like the Milky Way and their presence in some dwarf galaxies, today.
MBHs must have originated from {\it black hole seeds} that have grown in mass over time \cite{2002MNRAS.335..965Y}. These seeds may well have been born with masses straddling \cite{2017IJMPD..2630021M,2020ARA&A..58..257G} the two families of black holes, in what is currently an observational ``desert'', but that can be filled with different foremost experiments. 

The seeds can have primordial origin, generated by the collapse of high contrast density perturbations in the Early Universe, possibly as early as at the time of inflation, when the Universe experiences a brief phase of rapid exponential expansion. Inflation is also pivotal for generating structures: the small density perturbations generated by quantum fluctuations are accentuated by gravity  over time: since dark matter dominates the mass budget, initially these overdensities generate dark matter halos, within which baryons flow. When the baryonic gas in dark matter halos becomes sufficiently dense and cold, star formation initiates and galaxies are born. High initial overdensities eventually form massive galaxies and clusters of galaxies, while underdensities evolve into voids. Seeds can then be generated in galaxies, from the collapse of massive stars or from mergers of stars or stellar black holes. MBHs can reveal themselves directly only when they grow in mass: either by merging with another black hole, releasing gravitational waves (GWs), or by accreting gas or stars from their surroundings, when light is emitted as matter nears the horizon. Constraining the origins of MBHs requires therefore to trace their mass growth through the whole cosmic time theoretically and observationally in a truly multi-messenger effort.

\begin{tcolorbox}[title=Key points]
\footnotesize
\begin{itemize}

\item The discovery of quasars at cosmic distances and of giant dark massive objects in today's galaxies provide evidence of the ubiquity of massive black holes (MBHs).

\item Understanding the origins of MBHs goes hand in hand with understanding the origins of  structures inside the cosmic web. MBHs do not come to birth "massive" but must have grown by several orders of magnitude from "seed black holes". Gas accretion and black hole mergers are the drivers of their growth inside galaxies, but there are several bottlenecks in this journey. 

\item The origins of MBHs may be from exotic mechanisms or may well lie in known physics: particle, plasma and condensed matter physics, gravity and dynamics, but extrapolated to untested regimes. 

\item Studying the origins of MBHs is a multi-scale problem from the Schwarzschild radius to cosmological scales, from sub-second events to the age of the Universe. 

\item  Paths to seed formation and growth are not mutually exclusive, therefore constraints will come from a combination of observables: masses, spins, distances, spectra and light curves of populations of black holes. These indirect constraints can confirm that a given path exists, but cannot rule out the existence of other paths. A combination of electromagnetic and gravitational wave observations is the foreseen best strategy to constrain the origins of MBHs. 

\end{itemize}
\end{tcolorbox}
\normalsize

\section*{Channels of black hole seed formation}

The physical processes leading to the formation of  MBHs from ``gas clouds'' were first outlined by Martin Rees in 1978 \cite{1978IAUS...77..237R}. Most of the formation pathways \cite{2020ARA&A..58...27I} studied today are still based on these seminal ideas. A {\it seed}  is a black hole of yet unconstrained initial mass, in the range from a few hundreds to of order a million solar masses. We commonly refer as ``light'' the seeds with masses between $\sim 10^2\, \msun$ and a few $10^3\msun,$ and as ``heavy'' seeds those of $10^4\msun$ to $10^6\msun$. These middleweight black holes are {\it seeds} from which MBHs have grown if they experienced sufficient mass increase during the growth of dark matter halos and galaxies.
While the presence of MBHs in today's galaxies is ubiquitous, it may not have been the case at the time of seed formation.
With time, the assembly of cosmic structures modifies the {\it occupation fraction}, i.e., the fraction of galaxies and halos hosting an MBH, as they evolve: faster-growing halos have a chance to host an inherited seed as a consequence of a merger. This enhances the MBH occupation fraction over time. Here, we describe the astrophysical mechanisms of seed formation, discussing pros, cons, bottlenecks, with the warning that the prospected channels for forming seeds are not mutually exclusive. 

\subsection*{Channels linked to gas and stars}

Except for traces of lithium, no elements heavier than hydrogen and helium were synthesized  during Big Bang Nucleosynthesis, so that the first stellar objects originated within gas clouds of primordial composition cooling inside the lowest-mass, earliest forming dark matter halos. 
 Further generation stars formed in heavier halos, inside gas clouds enriched by trace amount of heavy elements released in supernova explosions. The lack of elements heavier than helium, called metals, provides favorable conditions for the birth of seeds. Because of fewer avenues for energy exchanges and spectral transitions, cooling is less efficient and gas has higher equilibrium temperatures. The concept of Jeans mass sets the scale of the collapsing gas clumps that give rise to stars: gravity overwhelms gas pressure in clumps with mass above the Jeans mass, which increases with temperature. Though this classic argument should be taken with caution \cite{1983ApJ...271..632P}, in the absence of metals, hotter Jeans unstable clouds fragment less so that stars form more massive. The same argument about spectral lines means also less effective radiation pressure, and at metallicities lower than 0.01 the solar value suppression of winds that carry away mass from stars leads to relic black holes with mass similar to that of the progenitor star \cite{Vink2001,2015MNRAS.451.4086S}. Low metallicity is therefore in principle doubly favorable as it leads to heavier stars and heavier black holes at fixed stellar mass.

 \subsubsection*{Single very/super massive stars }

The most common theories of seed formation are thus based on the collapse of stellar-like objects formed in metal-free and metal-poor environments \cite{2001ApJ...551L..27M}.  The first stars, dubbed {\it Population III} stars, originate in metal-free clouds hosted in halos with mass $\sim 10^6 \msun$ where the only coolant is hydrogen in its molecular state. Initially, Population III stars were thought to live short, solitary lives with one massive star of more than $10^2\msun$, forming in each dark matter halo \cite{2002Sci...295...93A}. Later theoretical studies found instead that several stars form per halo, with a broad  mass spectrum  from $\sim 10\msun$ up to $\sim 10^3\msun$  
\cite{2011ApJ...737...75G,2016ApJ...824..119H}, making their role as seeds more questionable, although we expect to find  Population~III relics in almost all galaxies \cite{2002ApJ...571...30S}. For instance in a galaxy like the Milky Way models of  Population~III formation \cite{2020ApJ...897...58T} predict about $3\times 10^4$ relics, but with an average mass of about $10 \msun$, depending however on the chosen mass distribution at birth of the stellar population.

An alternative pathway is the so-called {\it direct collapse} scenario where most of the gas in a halo of mass $\sim 10^8 \msun$ contracts coherently and forms a single  {\it supermassive star} (SMS) of $10^4-10^6 \msun$ that then collapses to a massive seed \cite{2003ApJ...596...34B,2006MNRAS.370..289B,2006MNRAS.371.1813L,2010MNRAS.402.1249S,Montero2012}, possibly via an intermediate phase as {\it quasistar} where the power comes from accretion onto a growing small black hole\cite{2008MNRAS.387.1649B} rather than nuclear burning. The key to reach such high masses is rapid growth of the protostar on timescales shorter than its thermal (Kelvin-Helmoltz) timescale guaranteeing that it remains cold at its surface \cite{2020SSRv..216...48H} and ultraviolet radiation that can ionize, heat and rarefy gas is absent, so that accretion can continue. 
These conditions are favored in halos whose gas is metal free and able to form atomic hydrogen, but no molecular hydrogen:  this triggers collapse of most of the gas in a short time.  Since molecular hydrogen usually forms first, the presence of external photo-dissociating radiation \cite{2013MNRAS.433.1607L,2014MNRAS.445.1056V} or mechanical energy sourced by rapid mass growth of the halo \cite{2019Natur.566...85W} is needed  thus requiring somewhat contrived circumstances. This makes the sites favorable for this process relatively rare \cite{2016MNRAS.463..529H}, with the predicted number density in the range $10^{-9}{-}10^{-2}\rm \, cMpc^{-3}$, depending on the assumptions made. To invoke {\it direct collapse} to explain $z>6$ quasars the average density of seeds is not a sufficient criterion:  the loci favorable for SMS formation are often in satellites of massive halos, only the few with short sinking time to the center of the main halo are viable  \cite{2021MNRAS.502..700C}.
Recent simulations have also shown that SMSs may not be as massive as initially predicted but reach only a few thousands of solar masses \cite{2020OJAp....3E..15R} because the environment is turbulent, and accretion on the protostar does not last long enough to create an SMS. Furthermore, the role of angular momentum on forming a single SMS versus fragmenting into multiple less massive stars has not been thoroughly explored yet. 

Alternatively, very rare major mergers of metal-enriched, very massive galaxies at redshift $z\sim 8$ have also been advocated to be site of SMS formation. Such collisions enhance temporarily the rate of central gas inflow by removing angular momentum via strong tidal torques, possibly preventing cooling in the center-most regions \cite{Mayer2015}, although no ab-initio cosmological simulation including all required physics has shown the occurrence of this process.

In summary, low mass seeds are common, but high mass seeds are rare and depend on processes that we have not observed yet.  

\subsubsection*{Runaway stellar mergers in young, dense clusters}
This channel foresees the birth of seeds from the collapse of very massive stars (VMSs) of $\sim 200-10^3 \, \msun$ forming in dense, metal-poor stellar clusters, which undergo core collapse on timescales shorter than the life of the most massive stars \cite{2004Natur.428..724P}.
  The dramatic, albeit temporary growth of the central density (up to values of $10^{5-7}\,\msun \,\rm pc^{-3}$) sparks violent few-body interactions among single massive stars (or even protostars) and stellar binaries, ending with the formation of a VMS through {\it runaway collisions} \cite{2004Natur.428..724P,Frietag2006,Mapelli2016}. The large cross section that binaries offer and the excitation of large eccentricities during the stellar  encounters lead to the physical collision of stars that repeats in a stochastic fashion until the encounter time becomes longer than  a few Myrs and the VMS collapses into a black hole of comparable mass  \cite{Mapelli2016,Reinoso2018}.
   VMSs can further form in gas-rich mildly polluted clouds via competitive accretion onto few central massive stars or stellar mergers \cite{Chon-Omukai2020,Boekholt2018,2020ApJ...892...36T}.
  
  Dense, young clusters as well as more massive, compact nuclear star clusters in high redshift galaxies appear to be ideal grounds to nurture VMSs  \cite{2008ApJ...686..801O,Devecchi2009,2015MNRAS.451.2352K,Yajima2016,Sakurai2017}, although the effect of mass loss at collision in determining the true mass of a VMS is not fully quantified. 
      Seeds with birth masses around  $10^2 \,\msun -10^3 \, \msun$  may start forming $\sim$ 300 Myrs after the Big Bang \cite{Devecchi2012}. Collisions between dark matter halos  of a few $10^5 \msun$ can also provide the conditions needed to form massive clusters with a VMS \cite{2015MNRAS.451.2352K}. Overall, the requirement to have mild metal pollution to spark fragmentation inside ultra-compact clusters where VMSs form, leaves a wider time window for seed formation to happen as long as the metallicity remains sufficiently low that stellar winds are weak and the collapse of the star results in a black hole of similar mass. Metallicity becomes unimportant if a stellar black hole grows into a seed through runaway tidal capture of stars \cite{2017MNRAS.467.4180S}, extending seed formation to lower redshifts.

\subsubsection*{Hierarchical black hole mergers }

Black holes relics of massive stars in stellar clusters of $10^4$ up to $10^7\,\msun$ may undergo repeated mergers to grow to the intermediate scale \cite{Giersz2015}. Owing to their larger mass compared to the mean mass of stars, they sink by dynamical friction at the center of the cluster where they pair via binary-single close dynamical encounters. Exchanging off stars, they build a nested core of binary and single black holes. In close, repeated encounters,  gravitational binding energy can be traded for kinetic energy, and such interactions can eject binaries before they  merge, thus aborting the process of growth \cite{Sigurdsson1993}. Their retention and merger rate depend on their mass at birth with the heaviest, with masses in excess of $50\msun$, being the most favored for the growth via repeated mergers with other black holes \cite{Miller2002}. A remaining obstacle to grow a seed by multiple generation mergers is the  GW-induced recoil that the coalescence product receives \cite{Lousto12}. Despite this, many studies  \cite{2019PhRvD.100d1301G,Antonini2019} show that the hierarchical growth is possible in nuclear star clusters
with escape velocities in excess of $10^2\,\rm km\,s^{-1}$.

Major gas inflows in metal-poor, dense star cluster can further help triggering a chain of multiple mergers as inflows deepen the gravitational potential well leading to the formation of seeds with masses up to $10^3\,\msun$ \cite{2011ApJ...740L..42D,2014MNRAS.442.3616L}. In dense, gas-rich nuclear star clusters a single stellar black hole can also grow to become a seed if in its motion  it captures low angular momentum gas so that quasi-spherical accretion can proceed unimpeded \cite{2014Sci...345.1330A}. 

These formation channels are only now receiving attention, following the discovery of GW190521\cite{Arcasedda21}. The appealing side of this process is of having a milder dependence on metallicity,
and therefore to possibly occur over longer cosmic times and not only in the first galaxies.
 The main uncertainty is in quantifying the importance of GW kicks in terminating the sequence of hierarchical growth and this depends on the depth of the cluster potential well and on the mass and spin distributions of the black holes in the binaries, which starts to be determined, but only for the low redshift GW sources.

\begin{tcolorbox}[title=Box 1: Summary of MBH formation processes]
\footnotesize
    \begin{wrapfigure}{r}{0.6\textwidth}
        \begin{center}
            \includegraphics[width=0.6\textwidth]{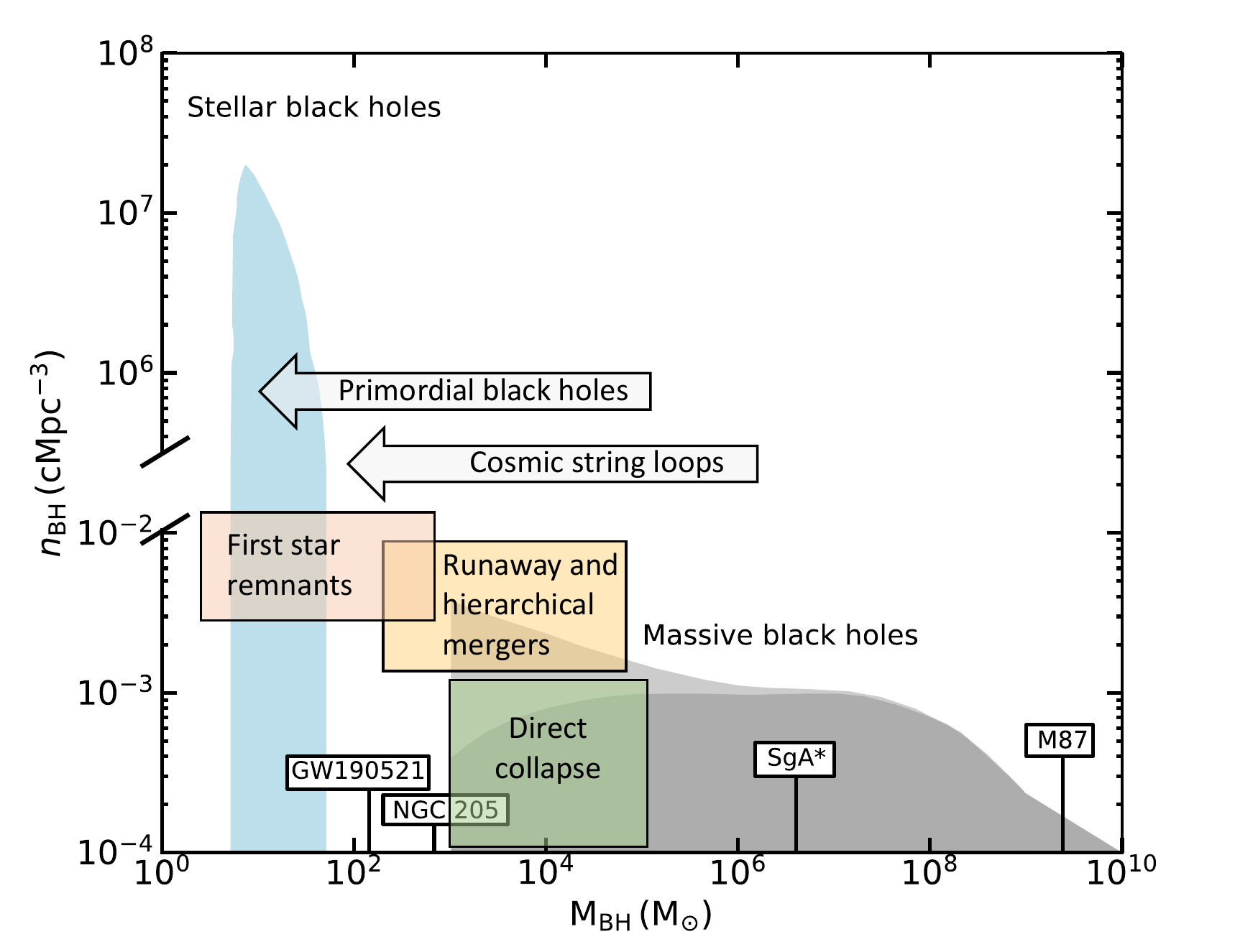}
        \end{center}
    \end{wrapfigure}
The origin of the MBHs embedded in the center of most galaxies is still unconstrained. They must have grown from seeds, i.e. black holes of unknown mass, but realistically from a few hundred to about a million solar masses. 
Because observationally unconstrained, numerous theoretical pathways to seed formation have been proposed, and they are likely to be not mutually exclusive. The mechanisms predict different seed mass, time of formation, environment, and abundance of MBHs. These mechanisms, or a combination of them, need to explain two key aspects of the observed MBH population: the observation of extremely luminous but rare \cite{2001AJ....122.2833F} ($n_{\rm BH}\sim 10^{-9} \, \rm  \, cMpc^{-3}$) quasars when the Universe was about 700~Myr and the abundance \cite{2020ARA&A..58..257G} of MBHs in the local Universe  ($n_{\rm BH}\sim 0.01-0.001\, \rm cMpc^{-3}$) in a variety of galaxies. We show a cartoon of the number density in stellar black holes and MBHs, with the shaded area highlighting the uncertainties for low-mass MBHs. We also qualitatively sketch the mass and density ranges for seeds from different channels, which populate the current observational ``desert'' between stellar and massive black holes. Seeds can have formed before or with galaxies. Among the first category, primordial black holes and cosmic string loops, set at arbitrary number densities in the figure.  The first galaxies appeared about 200-300 Myr after the Big Bang, and in potentially in a very large fraction of them, the most massive stars of the first stellar population could have collapsed onto seeds of $\sim 10^{2}\, \rm M_{\odot}$. Later in the history of the Universe, rarer supermassive stars could have collapsed onto seeds of $\sim 10^{4}-10^{6}\, \rm M_{\odot}$. Runaway accretion or stellar or black hole mergers in compact primeval star clusters could have triggered the formation of seeds with $\sim 10^{2}-10^{3}\, \rm M_{\odot}$. 
\end{tcolorbox}
\normalsize

\subsection*{Early Universe Cosmology Channels}
These are formation scenarios linked to the collapse of  regions in the Early Universe, before galaxy formation. In this case, galaxies form around black holes, and not black holes ``in'' galaxies: the presence of a black hole in an otherwise homogeneous Universe with small density fluctuations represents an overdensity that would have attracted dark matter (if black holes are not all of dark matter) and gas, thus creating sites for further galaxy formation. The black hole would then be located at the very center of the forming halo/galaxy and would naturally grow from accretion of the infalling material and mergers with other black holes, if they are clustered at formation. The dark matter cocoon enveloping the central black hole and attracting surrounding baryons in the deepening potential well would allow the seed to grow until energy injection from star formation and supernova explosions, or from accretion on the seed itself, stir the gas and modulate further growth. Despite these appealing features, there is no observational evidence of these processes and they often require ad-hoc assumptions, for which one has relative freedom exactly because of the lack of observational constraints.

\subsubsection*{Primordial black holes}
Primordial black holes can form in the presence of a high contrast density fluctuations, which collapse unto themselves.
They are not a generic prediction of inflationary cosmology, but can be generated under specific conditions \cite{2020ARNPS..7050520C}, for instance in an inflationary scenario with a peak or a very high-density tail in the power spectrum of curvature fluctuations, or if there is a sudden change in the plasma pressure, e.g., during reheating or phase transitions. 
The initial mass is generally that of the horizon at the time of collapse, and can therefore span a wide range, from significantly sub-solar to millions of solar masses depending on the specific mechanism triggering the collapse, e.g., from $\sim 1$~g just after inflation to around a solar mass at the  quantum-chromodynamic phase transition to $\sim 10^5 \,\msun$ at the time of $e^{+}e^{-}$ annihilation \cite{Garcia-Bellido:2019tvz}.  
Primordial black holes are a viable candidate as MBH seeds \cite{2001JETP...92..921R,2018MNRAS.478.3756C} if they have an overall low mass density, compared to the total matter density, in order to not violate existing constraints \cite{2008ApJ...680..829R,2020PhRvR...2b3204S,2020ARNPS..7050520C}. Given that MBHs represent about $10^{-5}-10^{-6}$ of the total matter density, and that this number has increased by several orders of magnitude as seeds became MBHs,  mass windows exist for primordial black holes to be seeds.  

\subsubsection*{Cosmic string loops}
Cosmic strings are topological defects that can form during phase transitions in the Early Universe both in quantum field theories and in string theory. When cosmic strings intersect and form loops, they can collapse into black holes \cite{1989PhLB..231..237H}  if the loops have sufficiently low angular momentum to contract within their masses' Schwarzschild radius or if they accrete enough dark matter to trigger collapse. 
GWs from oscillating string loops impose the strictest limits on the power spectrum generated by cosmic strings. 
Taking into account these bounds and further assuming that the size of loops depends uniquely on the time of formation of the loop and that one loop should seed one MBH per galaxy, the resulting mass and redshift windows of interest to explain MBHs can be calculated \cite{2015JCAP...06..007B}, resulting in seed masses up to $10^5 \,\msun$ at $z\sim 30$. 

\section*{From seeds to massive black holes}

\begin{figure}
    \centering
    \includegraphics[width=\columnwidth]{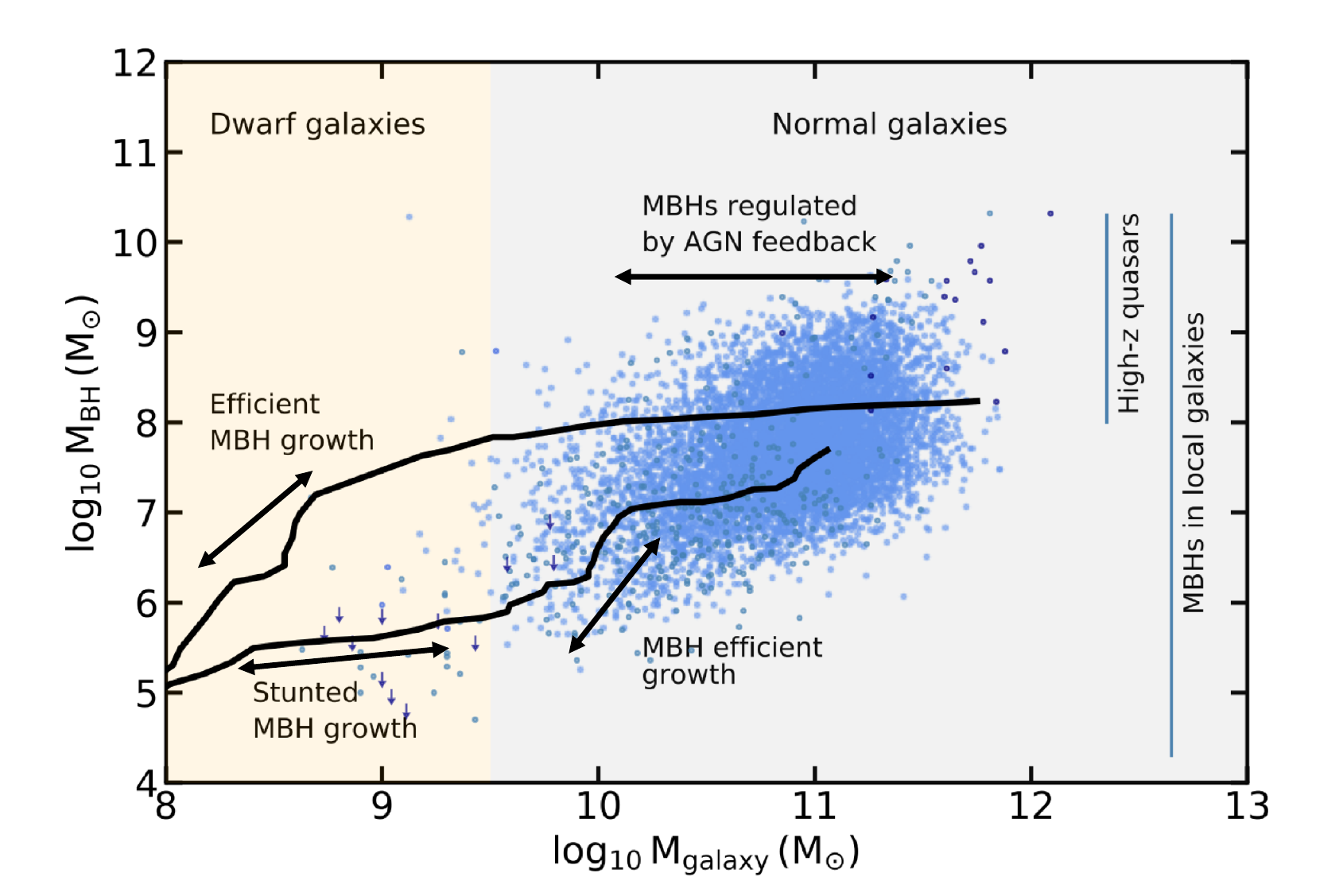}
    \caption{Growth of MBHs in galaxies:  MBH mass versus the stellar mass of the host galaxy. The blue cloud represents observational data \cite{2015ApJ...813...82R,2018ApJ...852..131B,2019MNRAS.487.3404B,2020ARA&A..58..257G}, while the curves are from simulations. Seeds must overcome hurdles to grow into MBHs, but when fully grown they regulate further growth through the energy and momentum they inject when feeding.}
    \label{fig3}
\end{figure}

The middleweight black holes whose formation channels we have sketched above are seeds for MBHs with masses millions to billions solar masses if they experienced sufficient mass increase during the growth of galaxies and dark matter halos (Fig.~\ref{fig3}). The growth of those seeds, ancestors of the MBHs powering quasars at redshift $z\sim 6-7$ is rapid, but it can be gradual for most of the population, and some of the black holes formed at early cosmic times may not germinate MBHs if conditions are unfavorable. 

Seeds of $10^2-10^4 \, \msun$ have to fight harder in order to develop into fully-fledged MBHs, as their gravitational sphere of influence is small compared to galactic dimensions and it is difficult to convey low angular momentum mass in the region where it can be captured by the seed.
If they form in low density environments and/or are too light to sink to the center of a galaxy potential well, their growth is immediately stunted \cite{2018MNRAS.480.3762S}. These conditions would not be suitable for efficient growth by gas accretion, neither by mergers with other seeds, and could lead to a significant population of low-mass wandering MBHs in galaxies. Furthermore, Population III relics have an additional complication since their progenitor stars produce copious ultraviolet radiation that sterilizes the environment and no immediate accretion is possible \cite{2007MNRAS.374.1557J,2018MNRAS.480.3762S}. 

While Population III relics may have difficulty in growing large by accretion despite their ubiquity in halos, the contingent conditions leading to the formation of an SMS are more exotic and such that the rate of direct collapse could be too low to explain the presence of MBHs in most galaxies. The dynamical channels based on runaway mergers of stars or black holes are elegant and based on known physics, but few investigations explored their role in building MBH populations, and primordial black holes and cosmic string loops are basically uncharted territory in MBH studies. 

\subsection*{Growth by accretion}

Most of the MBH growth in the Universe is sourced by accretion in the centers of galaxies \cite{2002MNRAS.335..965Y}, where  gas, having net rotation, approaches the last stable circular orbit gently spiralling down and forming an accretion disc.  Light is produced around the MBH, emerging from the accretion disc, the hot corona of tenuous hot plasma surrounding the disc and sometimes from a relativistic jet produced in the interaction of the magnetized plasma with the spinning MBH.  A small fraction of MBH growth can also be the product of accretion of gas stripped from stars passing very close to the MBH, thus creating transient sources called {\it tidal disruption events} \cite{1988Natur.333..523R}.  Accretion rate and conversion efficiency of gravitational energy into radiation determine the luminosity and the rate at which MBHs grow.

An important concept is that of Eddington luminosity: the maximal luminosity, linear in the MBH mass, resulting from the balance of the gravity force on infalling matter and  radiation pressure in spherical symmetry. This luminosity sets a maximal accretion rate, and a timescale for the $e$-fold increase of the MBH mass of $\sim 45$ Myr.  This concept is often used as a critical value for the maximal accretion rate onto an MBH: in reality accretion in discs can proceed above this critical value \cite{1982MitAG..57...27P}, as for instance has been observed in tidal disruption events, with the geometry of the disc changing and photons being trapped by the high densities, unable to diffuse out of the disc.

MBHs that are not (or weakly) accreting are called quiescent, with SgrA* in the Milky Way as a paramount example. Although they emit little light, quiescent MBHs can be studied through the orbital motion of stars and gas inside the region where dynamics and kinematics are controlled by the Keplerian potential of the MBH.

\subsubsection*{Growth hurdles: feedback and dynamics}

Many physical processes can slow down the growth of seeds by diminishing the MBH gas reservoir from which they accrete (Fig.~\ref{fig3}).
These processes are either related to the physics of the MBHs themselves, or to the physics of their host galaxies. 
AGN are able to release a large amount of energy on scales ranging from sub-pc to tens of kpc, potentially impacting their whole galaxies and beyond -- the most spectacular example are jets that can extend over physical scales of Mpc. The released energy or momentum 
couple to the gas in vicinity of the MBH.  This heats and/or depletes the gas reservoir \cite{1998A&A...331L...1S}, even preventing further gas infall \cite{2013MNRAS.428.2885D}, and can take several forms: thermal, mechanical through relativistic highly-collimated jets or non-relativistic non-collimated outflows, and electromagnetic (EM) radiation \cite{2012ARA&A..50..455F}. This is called AGN feedback, and  it is termed ``negative'' since it suppresses the source of energy, and can take place from birth in the case of high accretion rate seeds. An MBH would therefore cut off its own gas supply and stop accretion until the effect of feedback has vanished, and then a new cycle of accretion-feedback-starvation resumes. 

Explosions of supernovae near the MBH's gas reservoir can also suppress MBH accretion for similar physical reasons \cite{2015MNRAS.452.1502D,2017MNRAS.468.3935H,2017MNRAS.465...32B}. Supernova-driven winds can reach velocities of more than 200 $\rm{km \, s^{-1}}$ in the interstellar medium, enough to remove dense gas from low-mass galaxies with shallow gravitational potential wells. As a result, MBH growth could be delayed until their host galaxies are massive enough to retain the gas in the galaxy nuclear regions. This supernova feedback is thought to be the dominant hurdle in low-mass galaxies. 

Finally, MBHs accrete more efficiently when located in the center of galaxies, embedded in a dense gas reservoir, and have a hard time when displaced. This would be the case for MBHs too light to reach the potential well of their host galaxies \cite{2019MNRAS.482.2913B,2019MNRAS.486..101P}. These MBHs could wander in their galaxies for potentially a long time, and off-center MBHs have been detected in the local Universe \cite{2012Sci...337..554W,2015MNRAS.448.1893M,2020ApJ...888...36R,2020ApJ...898L..30M}. MBHs can also be kicked out of the center of galaxies after the coalescence with another MBH. The asymmetry of the GW emission from two merging MBHs, which depends on the mass ratio of the binary and on the spins magnitude and relative direction, can trigger the recoil of MBHs off the remnant galaxy center thus delaying its growth, or from the galaxy altogether, thus requiring the acquisition of a new seed. Triple MBH interactions can also displace them from galaxy centers.

\subsubsection*{Growth timescales for $z>6$ quasars}
An exceptional class of MBHs are those powering luminous quasars at $z>6$, currently up to $z=7.6$ \cite{2021ApJ...907L...1W}, when the Universe was less than one billion years old. These are among the most massive MBHs that we have discovered: they have masses of $10^8-10^{10} \,\msun$, comparable to those of the most massive quiescent MBHs in the local galaxies (Fig.~\ref{fig3}).  While rare, about 1 per Gpc$^3$, these objects pose a timing challenge as a seed of mass, e.g., $10^4 \,\msun$ born e.g., at $z=11$ or 0.42 Gyr after the Big Bang, must have grown continuously accreting at the Eddington luminosity with high radiative efficiency to reach $10^9\msun$ in the time elapsed between $z=11$ and 6.

This is a feat in light of the growth hurdles for MBHs in low mass galaxies, but if the galaxies hosting these distant quasars represent the descendants of remarkably luminous and massive $z=11$ galaxies like GN-z11\cite{2016ApJ...819..129O}, then their growth would be ``normal'' in ``exceptional'' galaxies that became massive enough to foster MBH growth very early on. In principle, such massive galaxies as GN-z11 anchor biased regions, and therefore we would expect an overdensity of galaxies around the quasars. 
Observationally there is no consensus on this \cite{uchiyama18,mignoli20}: several observed quasar environments on kpc to Mpc scales show no galaxy enhancement, and some could even be under-dense.  However, if we confirm that quasar fields are not statistically overdense, it would indicate a large variance in quasar environments \cite{2019MNRAS.489.1206H}, but also possibly the power of quasars to influence star formation in their own galaxy and even in nearby galaxies through their energy and momentum input \cite{2014MNRAS.439.2146C}. 

Solutions to the timing problem include more massive seeds, provided that they are not too rare, a significant contribution of MBH mergers to the mass budget or some phases of accretion at super-Eddington levels. MBH mergers, however, can have both a positive and negative impact on the mass growth because they can kick MBHs out of galaxies  \cite{2004ApJ...613...36H,Sijacki2009}.  The possibility of super-Eddington accretion is enticing, since it is a natural expectation of accretion disc physics \cite{1982MitAG..57...27P} and it has been predicted to occur in high-redshift galaxies under different conditions \cite{2005ApJ...633..624V,2016MNRAS.459.3738I,2014Sci...345.1330A}, including large reservoirs of gas and efficient transport of angular momentum from the galactic scale down to the black hole gravitational sphere of influence owing to turbulence \cite{2014ApJ...784L..38M} or angular momentum cancellation from multiple accreting streams \cite{2012MNRAS.423.3616D,2015MNRAS.453.1608C}, both more common at high redshift. The main uncertainty is for how long super-Eddington accretion can be sustained, since it is predicted to be accompanied by powerful jets that can destroy the flow of gas from the galaxy towards the MBH on small scales \cite{2019MNRAS.486.3892R} unless the jet can pierce through the whole gas distribution without affecting it, for instance if gas is confined in an equatorial disc down to the MBH accretion flow scale \cite{2020MNRAS.497..302T}. It is unclear if such axisymmetric conditions can be achieved taking in account the expected turbulent nature of high redshift galaxies.
\subsection*{Growth by mergers}

MBH mergers play a secondary role in the mass budget of MBHs, but they become important  when gas is not readily available \cite{2014MNRAS.440.1590D,2015ApJ...799..178K} or in galaxies with a rich merger history, and these conditions are generally met in low-redshift massive galaxies. The role of mergers in growing seeds at high redshift is still controversial: on the one hand  the merger rate of halos was higher at early times and the dynamical timescales were shorter because the Universe was denser, on the other hand the lightness of the seeds and the tumultuous evolution of high redshift galaxies makes it harder for seeds to find a companion and merge. 

General relativity predicts that accelerated masses in binaries emit GWs and coalescing stellar and massive black hole binaries are among the loudest sources of high and low frequency GWs, respectively \cite{2009LRR....12....2S}.  Here we briefly outline pathways for the formation of binaries conducive to mergers.  

\subsubsection*{In situ formation of black hole binaries}

Simulations of pristine star-forming clouds show that massive Population III stars form in binaries  \cite{Stacy2016,Sugimura2020}, but there is no proof that these binaries merge in situ shortly after their formation.  These mergers, when delayed in time, could contribute at percent level to the population detectable with advanced LIGO/Virgo \cite{2014MNRAS.442.2963K,2016MNRAS.460L..74H}.

As far as VMSs are concerned, numerical simulations of young star clusters suggest  that VMSs  acquire stellar or black hole companions, as natural outcome of the runaway process and that a fraction of these binaries have coalescence times shorter than the Hubble time \cite{Mapelli2016}.  Within the direct collapse scenario, general-relativistic numerical simulations show that differentially, rapidly rotating SMSs can split, during their collapse, into a black hole binary which merge shortly after if a bar-mode density perturbation is implanted in the SMS  \cite{2013PhRvL.111o1101R}. This may require fine tuned conditions since building a SMS by accretion requires low Keplerian velocities at the surface and thus slow rotation \cite{2019A&A...632L...2H}.  It has been further speculated that if mild fragmentation occurs around the SMS a companion star can form at close separations and create a binary \cite{2018MNRAS.479L..23H}.

\subsubsection*{Massive black hole binaries in galactic collisions}

Close pairs of MBHs are a natural outcome of galaxy collisions \cite{2014SSRv..183..189C,2019NewAR..8601525D} and can be discovered when the two MBHs are accreting. Dual AGN have been observed in interacting galaxies \cite{Liu2013} while MBHs bound in a Keplerian binary on sub-parsec scales are  challenging to observe but the number of candidate systems is increasing \cite{Eracleous2012,2021MNRAS.500.4025L}.

 To become sources of GWs, MBHs have to travel a long journey before reaching coalescence \cite{1980Natur.287..307B}, which occurs on a scale of a few micro-parsecs.  In galaxy mergers of comparable mass, dynamical friction against stars leads to the formation of a bound system, a binary still surrounded by stars and gas. The final crossing into the GW-driven regime then hinges on energy transfer through  scatterings of single stars impinging on the binary \cite{2015ApJ...810...49V}, and in the presence of a gaseous circum-binary disc through the interplay between gravitational and viscous torques\cite{2012A&A...545A.127R}.  In unequal mass galaxy mergers \cite{2009ApJ...696L..89C,2017ApJ...840...31D}, the tidally disrupted lighter galaxy may leave a wandering MBH or the binary evolution may stall until a third MBH comes into play to trigger a complex three-body dynamics which may end in the merger of the two closest MBHs \cite{2019MNRAS.486.4044B}.
In massive nuclear discs, outcome of galaxy mergers, the presence of giant gas clouds and of supernovae explosions \cite{2015MNRAS.453.3437L} that rarefy the medium lead to a broadening of the time elapsing between the galaxy and MBH mergers, and in some dwarf galaxies mergers can be delayed significantly due to their loose density structure \cite{2018ApJ...864L..19T}.

 At high-redshift, the dynamics of seeds inside forming  galaxies is key for determining their growth via accretion and  mergers: seeds need to remain anchored to the minimum of the gravitational potential well where densities are the highest. But, galaxies at these redshift are turbulent, clumpy, star-bursty and their morphology continues to change: thus they lack well defined stable centers. Under these circumstances, seeds have a complex erratic dynamics \cite{2019MNRAS.486..101P,2020MNRAS.498.3601B}.

\subsection*{Massive black holes and galaxies }

MBHs are generally found embedded in the center of most galaxies at the present time. 
In the local Universe, low-mass MBHs have been detected in dwarf galaxies \cite{2020ARA&A..58..257G}, and the heaviest MBHs are found in massive elliptical galaxies (Figure~\ref{fig3}). 
Some examples are the $5\times 10^4 \, \msun$ MBH in the dwarf galaxy RGG~118 \cite{2015ApJ...809L..14B}, Sagittarius A$^*$ in the Milky Way with a mass \cite{2008ApJ...689.1044G,2010RvMP...82.3121G} of $3.6\times 10^6 \msun$ and the $6.5\times 10^{9}\,\msun$ MBHs in M87 \cite{2019ApJ...875L...6E}. With time, we have accumulated evidence for a variety of empirical relations such as those between MBH mass and galaxy bulge luminosity, mass, stellar velocity dispersion  \cite{1998AJ....115.2285M,2000ApJ...539L...9F,2000ApJ...539L..13G,2009ApJ...698..198G}. Massive present-day galaxies assembled through accretion from cosmic filaments and galaxy mergers, and MBHs through accretion of gas in the galaxy and MBH mergers: but what sets the relative growth of one with respect to the other? Several aspects of galaxy evolution and of MBH assembly can shape such relations (and their scatter), making their interpretation a subject of intense debate.

At the MBH massive end, MBHs could impact the mass growth of their host galaxies and regulate themselves through AGN feedback, whose importance is expected to decrease towards low masses. In this regime galaxies could instead influence, potentially completely stunt, the growth of their MBHs through e.g., supernova feedback (Figure~\ref{fig3}).
Therefore such scaling relations and their scatter could be evidence for the complex co-evolution of MBHs and galaxies \cite{2011Sci...333..182H}, and help constrain MBH formation.
Whether and how these scaling relations evolve towards higher redshifts are  key questions, but selection effects and biases make such investigations very challenging.

\section*{Diagnostics on the origins}
The future is bright as planned and proposed facilities will target the origins of MBHs using light and GW as messengers \cite{2021NatRP...3..344B}. Seed formation channels are not mutually exclusive. Tests based on population studies can therefore identify the dominant channel, while tests based on individual ``special'' outlier sources can support one particular theory without necessarily ruling out the others.  To break degeneracies we need to combine observations of nearby galaxies with those of the farthest AGN. Traditional EM observations can give constraints on the broad population of accreting sources, through a combination of MBH mass and accretion rate, or on a smaller number of quiescent MBHs. GW observations yield directly the MBH masses, together with the distance, of merging MBHs. The key to unveil the majority of MBHs is the synergy between these different approaches. 
For instance, detecting the GW signal from a  binary  of  $\sim 100\msun$  at $z=10-20$ would prove that massive stars form at cosmic dawn but whether these are seeds requires additional proof, such as  detection of an accreting black hole of $10^3 \,\rm M_{\odot}$ at $z\sim 10$. This would be witnessing the growth of a light seed. Likewise, the GW detection of a $10^5\msun$ MBH at redshift $z>15$ would hint to direct collapse or Early Universe Cosmology Channels, but if we only detect merging MBHs of $\geqslant 10^{5}\, \rm M_{\odot}$, it does not rule out the existence of light seeds. Inefficient sinking of any type of seed to the galaxy center  could be responsible for the absence of GW detections. In that case, EM missions would be the only probes. On the other hand, if we mostly detects GWs from merging binaries of $<10^{5}\, \rm M_{\odot}$, identification of heavier accreting MBHs at $>10$ with EM telescopes would be required to not rule out heavy seed mechanism.

\subsection*{Local Universe}

\subsubsection*{Electromagnetic observations}

One possibility, albeit indirect, to obtain information on the origins of MBHs is  through ``archeology'' in local dwarf galaxies with stellar mass below $\sim 10^9 \,\msun$.  For this, theoretical predictions need to link the ``initial conditions'' to what we can observe today: how mass grows through accretion and mergers and the complex dynamics of sinking and ejections become key. Many studies of seeds are instead devoted only to the formation phase and do not evolve a population all the way to today, and even when they do they often include only one type of seeds.

The expectation is that measuring the masses of MBHs and their occupation fraction in the smallest dwarf galaxies are ways to infer which MBH formation channel is the most common \cite{2008MNRAS.383.1079V,2012NatCo...3.1304G}. However, many different seed models are predicting similar initial masses, $\sim 10^3 \,\msun$ (see Box~1), limiting the value of this probe, but the discovery of a $\sim 10^5 \,\msun$ MBH in a dwarf  galaxy with stellar mass $\sim 10^6 \,\msun$ -- that can be undoubtedly proved not to be the stripped core of an originally larger galaxy -- would hint to the existence of direct collapse.

 Furthermore MBH formation mechanisms predict that only a fraction of dwarf galaxies host MBHs, and this fraction depends on the specific mechanism. Finding MBHs in most dwarf galaxies, and especially those with stellar mass as low as $\sim 10^6 \,\msun$, would be a clear indication that MBH formation is common and therefore models that predict rare seeds, such as direct collapse or some flavors of primordial black holes, cannot be the dominant channel, as shown in Box~1. This is reflected in the shape and normalization of the MBH mass distribution, known as {\it mass function},  \cite{2010PhRvD..81j4014G}: if seeds are heavy the mass function would be truncated at higher masses than if seeds are light. The number density also reflects the frequency, or rarity of seeds:  to  explain today's MBH we need models that predict at least the observed number density.

Furthermore, finding MBHs in dwarf galaxies is observationally challenging: in galaxies with a shallow stellar distribution, MBHs may end up not being located in the galaxy center \cite{2019MNRAS.482.2913B}, making their detection more difficult. Success in finding MBHs in dwarf galaxies has been  obtained when selecting galaxies with dense stellar nuclei \cite{2019ApJ...872..104N}: the high stellar density makes it easier to to anchor the MBH in the center and to measure its mass through its dynamical signature on the stellar orbits. 

Relying on AGN in dwarf galaxies is hampered by the intrinsic faintness: 
since MBH mass decreases with galaxy mass, dwarf galaxies host low-mass MBHs, which when accreting  are faint AGN \cite{2013ApJ...775..116R}. This makes them hard to detect and disentangle from other sources that could have similar emission properties, typically binaries formed by a compact object and a star (X-ray binaries) or remnants of supernova explosions. To obtain the MBH occupation fraction a sample based on faint AGN must also be corrected including the MBHs that are quiescent and therefore missed by observations \cite{2015ApJ...799...98M}.

\begin{tcolorbox}[title=Box 2: Summary of the diagnostics for MBH formation/evolution]
\footnotesize
The quest to constrain the origins of MBHs has already started, and it will intensify in the future with the combination of several instruments, techniques, and diagnostics. Here the MBH mass-redshift plane is overlaid with the reach of various instruments: the signal-to-noise curves for GW facilities (for non-spinning binaries with mass ratio 0.5) as well as the approximate ranges for various telescopes and satellites. The GW antenna LISA in space and the 3G interferometers on Earth, such as Einstein Telescope (ET) \cite{2010CQGra..27s4002P}, will explore the mergers of black holes with masses of $10-10^{7}\, \rm M_{\odot}$. These instruments will survey the entire sky, in area and depth, from the surroundings of our Milky Way to the first hundreds of Myr of the Universe. Complementary EM missions will probe the population of accreting MBHs when seeds were still young (400-900 Myr). In optical/near-infrared Roman and Euclid will map the most luminous MBHs up to redshift $z\sim10$.  The James Webb Space Telescope will characterize high-redshift MBHs up to $z\sim 10$, from luminous quasars to potentially newly formed MBHs. Next-generation X-ray missions (Athena, LynX, AXIS) aim at finding MBHs close to the redshift of their formation. Extremely large telescopes (such as the ELT) on Earth, with mirrors up to $\sim 40$~m diameter, will enable measuring MBH masses in a variety of galaxies, including dwarfs, which are expected to have recorded key aspects of MBH formation. Tidal disruption events from transient surveys, such as LSST with the Rubin telescope, and extreme mass ratio inspirals from LISA will also extend the demography of low-mass MBHs.  Multi-band and multi-messenger (GW+EM) observations will help maximize the scientific return. Credits: Valiante et al. 2021 and NASA/WMAP Science Team.
        \begin{center}
            \includegraphics[width=\textwidth]{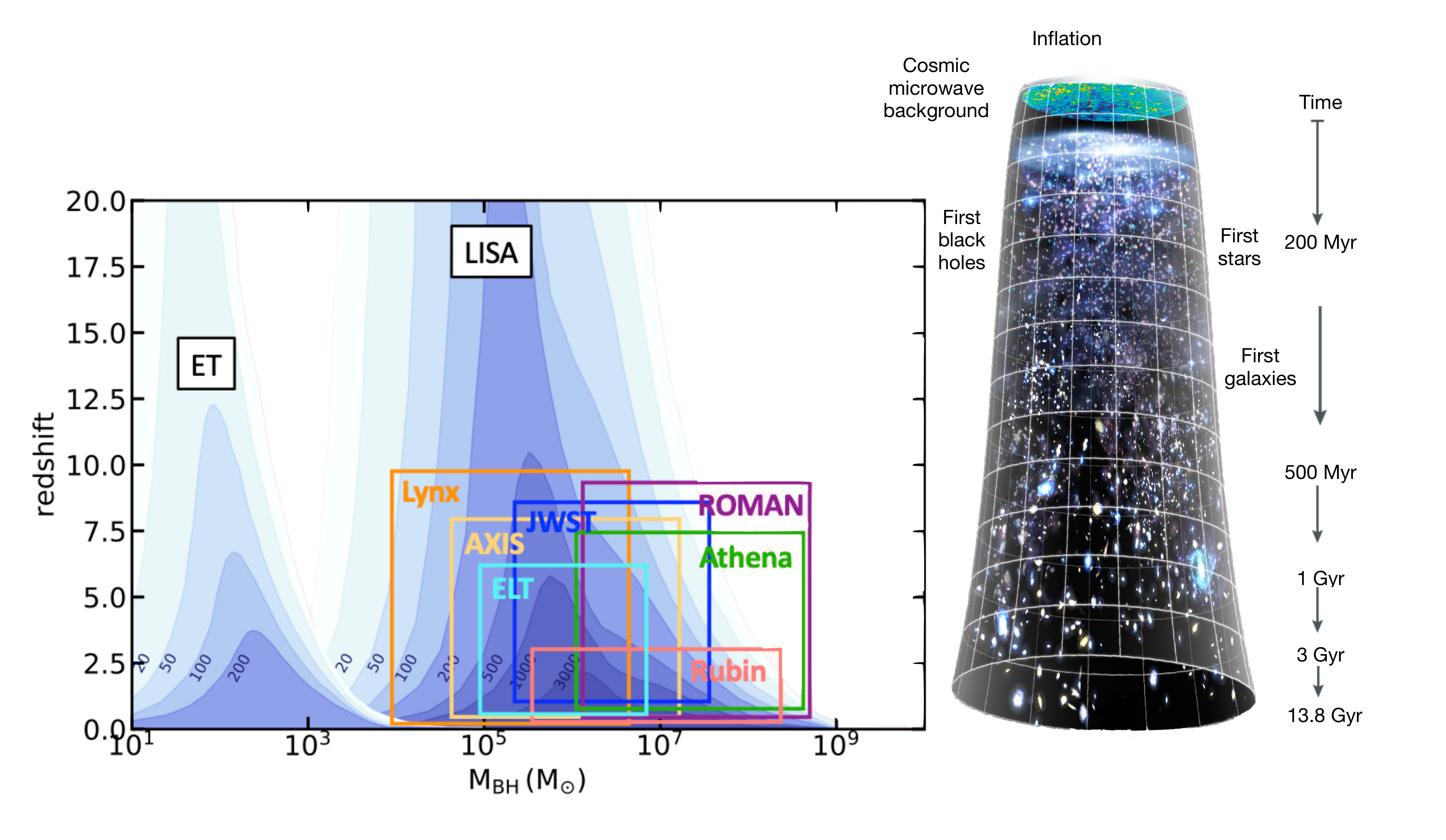}
        \end{center}
\end{tcolorbox}
\normalsize

Tidal disruption events are a very promising route to connect MBHs to their origins, being sensitive to the occupation fraction \cite{2016MNRAS.455..859S}. Tidal disruption events occurring around MBHs lighter than about $10^8\msun$ generally trigger accretion episodes around or even above the Eddington luminosity, thus making these MBHs as bright as they can be for a few months, until their luminosity decays. Surveys with capabilities to detect transient sources, such as the Zwicky Transient Facility, the Legacy Survey of Space and Time (LSST) with the Rubin Telescope and eRosita will greatly increase the number of known tidal disruption events, which can then be followed up to estimate the MBH and galaxy properties, and enlarge the census of MBHs in dwarf galaxies to constrain the occupation fraction and the asymptotic MBH mass in dwarfs. The MBH masses can be estimated from the width of emission lines, using virial arguments, in the case of active MBHs \cite{2004ApJ...607...90B} or measured through the kinematics of stars and gas, which have been proven successful even in the dwarf galaxy regime \cite{2019ApJ...872..104N,2020MNRAS.496.4061D}.

Discovering black holes of $\sim 10^3 \msun$ in the star clusters of nearby (star forming) galaxies would provide compelling evidence that dynamical interactions, runaway or hierarchical, leading to black holes of $10^2-10^3 \msun$ occur in Nature.  Metallicity, age and compactness of the cluster could help refine which of the two avenues,  runaway versus hierarchical merger, is the most likely.

\subsubsection*{Gravitational Wave observations}

The recent discovery of GW190521 \cite{2020ApJ...900L..13A}, the binary which produced the heaviest stellar-mass black hole ever observed of $142^{+28}_{-16}\msun $ has already sparked a lively discussion on its origin, whether it formed in a field binary or via stellar collisions or hierarchical mergers of lower-mass black holes in star cluster, i.e.  via the same mechanisms proposed for the formation of seeds \cite{2020ApJ...900L..13A}. While the GW antennas on Earth are already uncovering  systems that could represent the low-redshift analogs of seeds, LISA \cite{2017arXiv170200786A} can contribute to our understanding of the origins of MBHs with Extreme Mass Ratio Inspirals (EMRIs). EMRIs occur when a compact object, such as a stellar-mass black hole, is captured on a relativistic orbit by an MBH and slowly inspirals solely by emission of 
GWs, covering $10^{4-5}$ orbits until it crosses the event horizon. EMRIs are of pivotal importance to map the spacetime and test the nature of gravity, and the MBH mass and spin can be measured with exquisite precision \cite{2017PhRvD..95j3012B}. LISA can detect EMRIs around MBHs with mass $\sim 10^4-10^7 \,\msun$, with systems of mass $\sim 10^6\,\msun$ detectable up to $z \sim 4$ \cite{2017PhRvD..95j3012B} .  If LISA discovers enough EMRIs \cite{2017PhRvD..95j3012B}, we can probe the MBH mass function \cite{2010PhRvD..81j4014G} at low mass, which, as discussed in the previous section, is a sensitive probe of seed formation channels. The combination of the information from EMRIs and EM searches in dwarf galaxies represents an excellent synergy to constrain seed formation. The rate of EMRIs is currently unconstrained: different mechanisms, each with large uncertainties, can lead to their formation \cite{2018LRR....21....4A}, from the classic mechanism of stellar dynamical relaxation around an MBH \cite{2011PhRvD..84d4024M}, to breaking of binaries composed of a stellar black hole and a star by the tidal field of the MBH \cite{2005ApJ...631L.117M}, to migration in the accretion discs of AGN \cite{2011PhRvD..84b4032K}. 

\begin{tcolorbox}[title=Box 3: Massive black holes and dark matter]
\footnotesize
Black holes and dark matter \cite{2019arXiv190710610B} share the characteristics of being somber: most of their interactions are gravitational. Early on, dark matter was proposed to be simply faint baryonic compact objects. Constraints from the cosmic microwave background and primordial nucleosynthesis demonstrated however that dark matter cannot be composed of baryonic matter.  This line of reasoning leaves open the possibility that primordial black holes, formed before nucleosynthesis, constitute all or a fraction of dark matter. This subject has been explored with varied levels of interest and popularity over the last 40 years \cite{2020ARNPS..7050520C}. A large fraction of the parameter space is ruled out, for instance Hawking radiation provides a lower bound to the the mass of primordial black holes that have survived to today, while a range of observations ranging from microlensing, to LIGO/Virgo mergers, to galactic dynamics to the cosmic microwave background leave narrow windows for primordial black holes to constitute all of dark matter. No matter what dark matter is, it interacts gravitationally with black holes, and in stable long-lived conditions dark matter would accumulate in a density spike around an MBH in the same way that stars and compact objects do. These spikes are fragile to perturbations, but if dark matter can annihilate producing $\gamma$-rays, annihilations would be highly enhanced in these high-density spikes \cite{1999PhRvL..83.1719G}. Dark matter, and these spikes, would also alter the dynamics of gravitational wave-emitting binary black holes \cite{2013ApJ...774...48M,2020PhRvD.102h3006K}. If dark matter is composed of ultralight bosons, clouds form around black holes, extracting angular momentum from the holes, thus spinning them down \cite{2011PhRvD..83d4026A} and producing quasi-continuous GW signals and a stochastic background that one could search for with LISA \cite{2017PhRvL.119m1101B}. Finally, in the case of self-interacting dark matter, the shallower density profiles of dark matter halos (compared to standard cold dark matter) lead to less effective seed formation, growth and dynamical evolution \cite{2017MNRAS.469.2845D,2021MNRAS.500.2177C}. 
\end{tcolorbox}
\normalsize

\subsection*{High-redshift Universe}

\subsubsection*{Electromagnetic observations}

Beyond $z\geqslant 6$, our current facilities only allow us to detect the EM signal of the most massive, $M_{\rm BH}\geqslant 10^{8} \,\msun$, and luminous quasars with bolometric luminosity of $L_{\rm bol}\geqslant 10^{46}\, \rm erg\,s^{-1}$.
These rare objects increased their mass by many orders of magnitude since birth, erasing the memory of the seeding and growth history. Thus, they provide us with degenerate constraints on their initial seed mass and in addition they represent only the tip of the iceberg of the whole MBH population.  Addressing the origins of MBHs requires to be able to find seed-mass MBHs close to the redshifts of their formation. Hard X-rays can penetrate through the dense gas clouds in the vicinity of the MBH, while optical radiation cannot. Thus, next-generation X-ray missions such as Athena \cite{2013arXiv1306.2307N}, or concepts as Lynx \cite{2018arXiv180909642T} and AXIS \cite{2018SPIE10699E..29M} aim at a substantial increase in the sensitivity, in order to have the ability to detect young accreting MBHs. Detecting and measuring the luminosity, and counting MBHs to build a {\it luminosity function} is not enough though: this will only constrain a combination of MBH parameters: seed mass distribution, accretion distribution, and probability for galaxies to host a MBH.

The synergy between these X-ray missions and the near-future deep imaging surveys of the James Webb (JWST, near, mid-infrared) and Roman (optical-near-infrared) space telescopes will help breaking these degeneracies. Different MBH formation mechanisms yield a different connection between the MBHs and their host galaxies. Small $M_{\rm BH}/M_{\star}$ ratios are favoured by mechanisms forming seeds as the remnants of Population III stars, as runaway stellar mergers in compact nuclear clusters, or as runaway black hole mergers. In the direct collapse model, instead, newly formed MBHs are predicted to be over-massive compared to their host galaxies, with large $M_{\rm BH}/M_{\star}$ ratios of about unity \cite{2013MNRAS.432.3438A}. Therefore, the combination of MBH and galaxy properties could provide us with direct high-redshift constraints on MBH origins. The spectral energy distribution of high-redshift systems, observed with JWST, could also help us to distinguish between different seeding mechanisms \cite{2017ApJ...838..117N,2018MNRAS.476..407V}.  While in the case of light seeds the MBH emission would be outshined by the stellar component of the host galaxies and missed, direct collapse MBHs could outshine their galaxies in the infra-red, and would be detectable. Finally, within a narrow mass range SMSs could explode \cite{Montero2012} rather than collapse into black holes: by extrapolation, detection of such hypernovae could be a proof of the existence of SMSs outside this interval.

\subsubsection*{Gravitational Wave observations}

A network of next-generation terrestrial interferometers (3G) \cite{2010CQGra..27s4002P,2019BAAS...51g..35R}, and LISA from space will probe an immense volume of the Universe and of the MBH parameter space as their horizons extend  up to $z\sim 20-30$ and even beyond. This will let us probe black hole masses from  $100\msun$ up to $10^7\msun$. 3G detectors with sensitivity from a few Hz to a few kHz  have the unique capability of discovering the earliest binary light seeds forming in the Universe, and also those that failed to grow, the starved seeds that might be still present in galaxies at lower redshift \cite{2021MNRAS.500.4095V}. 
LISA, as all sky monitor sensitive to frequencies between $10^{-4}-10^{-1}$ Hz, has the potential to detect heavy seeds or light seeds in their transit to become massive. But, disentangling between the two is not easy. Binaries of $10^5\msun$  at $z\sim 10$ could either result from grown light seeds, or from just formed heavy seeds, or from both. 
If enough coalescences are detected, LISA can break this degeneracy between different evolution pathways \cite{2011PhRvD..83d4036S} by matching the information on the masses, mass ratios, spin and redshift distributions extracted from the signals with theoretical distributions. 3G detectors can further help in anchoring these physical models.  Having full view of the GW sky, encompassing also the deciHz  window, will allow us to follow the tracks travelled  by light seeds in their early growth. There have been a number of proposals for instruments that will probe this intermediate frequency range, for example DECIGO \cite {2009JPhCS.154a2040S}.
Lastly, robust identification of a merger at redshifts before stars are believed to have formed would provide concrete evidence for the existence of primordial black holes.

\begin{tcolorbox}[title=Box 4: Penrose's concept of black hole formation]
\footnotesize
Uncovering the origins of MBHs is an intellectual {\it voyage} into one of the most fascinating  discoveries of general relativity: the existence of absolute event horizons, responsible for the ``blackness'' of the ``holes'' emerging when gravitational collapse ``passes to a point of no return''. Penrose introduced  powerful concepts: that of {\it trapped surface}, formalizing the condition of light-ray convergence in collapsing matter,  and of {\it singularity} of spacetime that he characterized through the notion of incomplete causal geodesics. 
In his endless effort to understand the causal and global structure of spacetime, Penrose gave us the {\it weak cosmic censorship} stating that singularities are inaccessible to external observers, covered by the event horizon. 

Penrose gives an illuminating description of gravitational collapse to a black hole \cite{Penrose69}: {\it A body, or collection of bodies, collapses down to a size comparable to its Schwarzschild radius, after which a trapped surface can be found in the region surrounding the matter. Some way outside the trapped surface region is a surface which will ultimately be the absolute event horizon. But at present, this surface is still expanding somewhat. Its exact location is a complicated affair and it depends on how much more matter (or radiation) ultimately falls in. We assume only a finite amount falls. Then the expansion of the absolute event horizon gradually slows down to stationarity. Ultimately the field settles down to becoming a Kerr solution. }

Nature seems to have brought all of this into physical reality. This is testified by the discovery, through precision stellar-dynamical measurements by R. Genzel and A. Ghez, of a massive dark object in the midst of our Milky Way Galaxy, for which the MBH hypothesis appears the most natural, physical explanation. The Event Horizon Telescope has also given us a tantalizing image of light close to the event horizon of the MBH in M87 \cite{2019ApJ...875L...1E}. 
\end{tcolorbox}
\normalsize

\section*{Conclusions and perspectives: }

Black holes have the beauty of being the simplest, most elementary objects in the Universe. They are however involved in complex physical processes once they interact with their environment. They are powerful engines, being bright sources in the EM and GW skies. Our astrophysical understanding of black holes is rooted in the recognition that relatively massive stars collapse into neutron stars, and that the massive ones collapse into stellar black holes, since there is an upper bound to the mass of a neutron star. The physics leading to the formation of stellar black holes stands on solid ground, despite uncertainties on the state of matter above nuclear densities.  By contrast, the physics leading to the formation of MBHs is not known.  MBHs may originate from primordial black holes,  or from relatively light black holes, both rearranged and transformed into giants by accretion and mergers. SMSs would die as heavy seeds in response to general relativity instabilities excited inside radiation dominated, low density plasmas whose micro-physical state is well known.  Alas, we do not have observational evidence yet of the collapse ``to the point of no return'' of such massive stars. 

What we know is that upcoming ground- or space-based telescopes aim at constraining the AGN population at high redshift, potentially down to low-mass MBHs close to the seeding mass, while the detections of MBHs in local dwarf galaxies can provide indirect constraints by analogy with high-redshift systems. These EM observations complement GWs that will constrain the merging MBH population ($10^2-10^6 \, \msun$), active or quiescent, but are also limited to the binary population.
We are entering the era of ``precision GW astrophysics'' detecting signals from coalescing black holes on {\it all} mass scales to witness the emergence of astrophysical event horizons in dynamical spacetimes, and signals from  EMRIs, the sources that will let us  probe the global, causal structure of stationary spacetimes around central dark massive objects.

There are various facets to the origins of MBHs and we have to consider them together since they are interrelated: we need to be open-minded about connecting different fields of research  -- don't look under the lamppost, there is another lamppost not far away.

\section*{Glossary terms}
\begin{itemize}
\item Active Galactic Nuclei (AGN) and quasars: sources powered by an accreting MBH. Quasars are the most luminous among AGN. 
\item Eddington luminosity: maximal luminosity above which radiation pressure on electrons overcomes gravity on the infalling matter, under the assumption of spherical symmetry. 

\item Redshift: short for ``cosmological redshift'' in this review, and used as indicator for distance and cosmic time. Given a cosmological model there is unique relation between the redshift of a source and its distance from us, as well as the age of the Universe at that redshift.
\item Parsec, pc: a unit of length used in this review that corresponds to $3.0857\times 10^{18}$~cm. 
\item Solar mass, $\msun$: a unit of mass that corresponds to $1.98847\times 10^{33}$~g. 
\item Compact objects: are the relics of stars and comprise white dwarfs, neutron stars and stellar black holes.
\item Comoving and proper distances: comoving distances -- for which we prefactor a letter ``c'', e.g., cMpc -- are independent of cosmic expansion, while proper distances account for that, so that proper distances decrease at earlier cosmic times. 
\item Feedback: physical processes in which the energy/momentum output of a system (or a fraction of the output) returns to or impacts the system's input.
\item Metallicity: metallicity is the sum of the mass fraction of all the elements present in the system  heavier than hydrogen and helium. For metal-enriched systems, the Sun is often used as a unit of measure for metallicity, with $Z_{\odot}=0.012$. 
\item Radiative efficiency: $\varepsilon$ is the efficiency at which gravitational energy is converted into radiation. It establishes the link between the accretion luminosity $L$ and mass accretion rate $\dot M$: $L=\varepsilon{\dot M} c^2.$  In geometrically thin, optically thick accretion discs around black holes, $\varepsilon\sim 0.06-0.32$ in dependence of the spin, with $\varepsilon\sim 0.1,$ used as reference value. $\varepsilon$ can be lower depending on the geometry of the flow. $(1-\varepsilon) {\dot M}$ gives the mass growth rate of a MBH.

\end{itemize}

\bibliography{references}

\section*{Acknowledgements (optional)}
The authors thank Fabio Antonini, Robert Brandenberger, Gianfranco Bertone, Jonathan Gair, Jenny Greene, Kohei Inayoshi, Michela Mapelli, Priyamvada Natarajana and Paolo Pani for comments on the manuscript and T. Hartwig for estimating the number of Population~III relics for this review. 

\section*{Author contributions}
The authors contributed equally to all aspects of the article. 

\section*{Competing interests}
The authors declare no competing interests.

\section*{Publisher's note}
Springer Nature remains neutral with regard to jurisdictional claims in published maps and institutional affiliations.

\end{document}